\documentstyle[psfig,a4wide,pre,aps]{revtex} 

\def \av#1{\langle #1 \rangle}

\draft
\psfigurepath{figs}
\begin{document}

\title{Sliding blocks with random friction and absorbing random walks}
\author{A.R. Lima$^1$\cite{arlima}, C.F. Moukarzel$^1$, I. Grosse$^2$ and
  T.J.P.  Penna$^1$}
\address{ (1) Instituto de F\'{\i}sica, Universidade Federal
  Fluminense\\Av.  Litor\^anea  24210-340,  Niter\'oi, RJ, Brazil\\
  (2) Center for Polymer Studies, Boston University\\590 Commonwealth Ave.
  02215, Boston, MA, USA }
\date{\today}
\maketitle

\begin{abstract}
  With the purpose of explaining recent experimental findings, we study the
  distribution $A(\lambda)$ of distances $\lambda$ traversed by a block that
  slides on an inclined plane and stops due to friction. A simple model in
  which the friction coefficient $\mu$ is a random function of position is
  considered. The problem of finding $A(\lambda)$ is equivalent to a
  First-Passage-Time problem for a one-dimensional random walk with nonzero
  drift, whose exact solution is well-known. From the exact solution of this
  problem we conclude that: a) for inclination angles $\theta$ less than
  $\theta_c=\tan(\av{\mu})$ the average traversed distance $\av{\lambda}$ is
  finite, and diverges when $\theta \to \theta_c^{-}$ as $\av{\lambda} \sim
  (\theta_c-\theta)^{-1}$; b) at the critical angle a power-law distribution
  of slidings is obtained: $A(\lambda) \sim \lambda^{-3/2}$.  Our analytical
  results are confirmed by numerical simulation, and are in partial agreement
  with the reported experimental results. We discuss the possible reasons for
  the remaining discrepancies.
\end{abstract}
\section{ Introduction }
\label{sec:intro}

Friction between solid surfaces is present in everyday life. One of
the first experimental studies on friction was done by Leonardo da
Vinci. His studies were rediscovered and announced by Amonton de la
Hire in 1699 in the form of two laws: friction forces are a)
independent of the size of the surfaces in contact; and, b)
proportional to the normal load. The proportionality coefficient $\mu$
is the friction coefficient, and depends on the material.  The
influence of velocity was later studied by Coulomb, who discussed the
difference between static and dynamic friction.  Since then many
studies of friction have been conducted, which has revealed the
complexity of friction related
phenomena~\cite{examples,wolf96,tabor50,bbgr96,webman99,caroli96,brito95}.
The study of friction has been the subject of renewed interest lately
due to its relevance in the behavior of granular
materials~\cite{wolf96}.

Due to surface roughness, the interface between two solids put in contact can
be thought to consist of many points, rather than a continuous
region~\cite{tabor50}.  These contact points define a two-dimensional random
set called ``multicontact interface''. A basic setup for experiments on
multicontact interfaces consists of a slider of mass $M$ pulled by a spring
with effective stiffness $K$ (that could represent the bulk elasticity of the
solid), at a driving velocity $v$~\cite{bbgr96}. Depending on the parameters
$K,v$ and $M$, the sliding motion can have different regimes, including an
oscillating ``stick-slip'' instability.  Moreover, the friction coefficient is
found to depend not only on these three parameters but also on a variety of
other factors such as contact stiffness, creep aging and velocity weakening of
the contacts, that lead to a dependence not only on the instantaneous-velocity
but also on the sliding history~\cite{bbgr96,webman99}.  Therefore, the
friction force seems to be both state- and rate-dependent.  A phenomenological
derivation of the friction force that reproduces some aspects of the
experimental data was proposed by Caroli and Velicky~\cite{caroli96}.

Here, we focus on the random character of the multicontact interface, and show
that a simple model whose only ingredient is a randomly varying friction
coefficient can explain recent experimental findings. We consider in
particular the dynamics of a sliding block on an inclined plane. This problem
has been recently revisited by Brito and Gomes (BG)~\cite{brito95}, who report
unexpected results.  In their experimental setup, a block rests on a plane
which makes an angle $\theta$ with the horizontal, where $\theta$ is close to
but smaller than $\theta_c$, the critical angle for dynamic friction. The
block is set in motion by the impact of a hammer at the base of the inclined
plane.  A ``sliding'' is so produced, and the block stops after traversing a
distance $\lambda$.  Measuring the distribution $N(\lambda)$ of slidings with
length larger than $\lambda$, these authors find that, for $\theta$ close to
$\theta_c$, $N(\lambda) \sim \lambda^{-\delta}$.  The exponent $\delta$ is
$\approx 1/2$ and does not seem to depend on the type of material that makes
the block. Further exponents can be in principle defined, such as the one
describing the divergence of the mean sliding length $\av{\lambda} \sim
(\theta_c - \theta)^{-\tau_1}$ as $\theta \to \theta_c^{-}$. Brito and Gomes
report $\tau_1 \approx 0.23$~\cite{brito95}.

In this work we introduce a model that uses a simple expression~\cite{tabor50}
for the friction force and provides a microscopic explanation for most of the
findings of Brito and Gomes. We assume that friction is due to the existence
of random contact points between the surfaces, therefore the friction
coefficient is a rapidly varying function $\mu(\ell)$ of the block position
$\ell$ on the plane. A fundamental hypothesis, which makes this model exactly
solvable, is that the distribution of contact points is {\em uncorrelated} on
the length-scales of interest. We focus here on the simplest realization of
the model, where no other features such as velocity-dependent forces are
included. This model has been studied numerically previously~\cite{lima98}. We
show here that a closed analytical solution can be obtained by mapping this
problem onto a First-Passage-Time problem.

This paper is structured as follows: in Section \ref{sec:model} our model is
described and some numerical results are presented. In Section
\ref{sec:theory} it is shown that this system is equivalent to a random walk
with an absorbing barrier, and an exact solution is derived for the
distribution of slidings. Also in this section a comparison is made between
numerical, analytical and experimental results. Section \ref{sec:conclusions}
contains a short discussion of our results.

\section{The Model}
\label{sec:model}

\begin{figure}[thpb]
  \centerline{\psfig{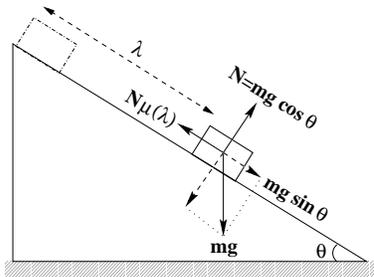}}
\caption{{}
Schematic representation for the block sliding on a chute.
  $\lambda$ is the displacement from the initial block position. The friction
  force depends on the block's position.}
\label{fig:1}
\end{figure}

Consider a block of mass $m$ on a plane making an angle $\theta$ with the
horizontal, and assume that at time $t=0$ the block is set in motion with
velocity $v_0$, i.e. with kinetic energy $K_0=m v_0^{2}/2$. Let $\ell$ be the
distance traversed by the block from its starting position, measured along the
plane, and $K(\ell)$ its kinetic energy. Since the friction force opposing the
movement is $mg \mu(\ell) \cos{\theta}$, and the parallel component of the
gravitational force is $mg \sin{\theta}$ (see Fig.~\ref{fig:1}), energy
balance implies
\begin{equation}
\label{eq:mov}
dK + \{ mg \mu(\ell) \cos{\theta} - mg \sin{\theta}\} d\ell = 0
\end{equation}
We rewrite this in terms of the reduced kinetic energy $k(\ell) = K(\ell) / mg
\cos{\theta}$ as
\begin{equation}
\frac{\partial k(\ell)}{\partial \ell} =  \tan{\theta} - \mu(\ell) 
\label{eq:redk}
\end{equation}
This equation can be integrated until the kinetic energy becomes zero. This
defines the ``avalanche size'', or stopping distance $\lambda$.  If $\mu(\ell)
= C$ independent of $\ell$, one has that \hbox{$\lambda_C =
  v_0^2/2g\cos\theta(C-\tan\theta)$}.  This does not in general agree with
experimental results \cite{brito95}, which show a broad distribution of
stopping distances. One could argue that in the experiments of BG, $v_0$ is
randomly distributed and thus $\lambda$ must show a distribution with a finite
width as well. But this sort of randomness cannot give rise to a power-law
distribution of stopping distances as observed in experiments, unless $v_0$
itself is power-law distributed, which doesn't seem to be easily justified.

Because of the random character of the multicontact interface, it is on the
other hand physically reasonable to assume that the coefficient of friction is
not constant but changes randomly from point to point.  In this case the
stopping length $\lambda$ becomes a stochastic variable, and we are interested
here in calculating the probability $A(\lambda)$ for the block to stop at a
given position $\lambda$. We will show that under certain circumstances (e.g.
close to the critical angle), fluctuations in the friction coefficient can
have important observable consequences, and in particular that such
fluctuations give rise to a power-law distribution of stopping distances.

For simplicity we assume $\mu(\ell)$ to be an uncorrelated random function of
position, i.e.
\begin{eqnarray}
\mu(\ell)\quad &=&\overline \mu - \eta(\ell) \qquad \hbox{where}\\
\av{\eta(\ell)} &=&0 \nonumber \\
\av{\eta(\ell) \eta(\ell')}&=&\sigma^2 \delta(\ell-\ell') \nonumber
\end{eqnarray}
So that (\ref{eq:redk}) now reads
\begin{equation}
\frac{\partial k(\ell)}{\partial \ell} =  V + \eta(\ell) 
\label{eq:reduced}
\end{equation}
where $V=\tan{\theta} - \overline \mu$ is the mean drift, and $\eta(\ell)$ is
a noise term. If the mean drift $V$ is positive, clearly there will be a
finite probability for the block never to stop. For $V<0$ on the other hand
the block always stops.

This problem can be easily implemented numerically~\cite{lima98}. In our
numerical implementation both the block and the plane surfaces are represented
by finite sequences of 0s and 1s, each bit corresponding to a small region of
length $a$. If a given region of the surface is ``prominent'', the
corresponding bit is set to one.  Similarly if that region is ``deep'', the
corresponding bit is set to 0. Thus the profile of these surfaces is
represented by strings of bits which are set to one with probability $C_p$ and
$C_b$ for the plane and block respectively. One says that the block and plane
are in contact at a given point whenever both the plane bit and the block bit
that sits on top of it are set to one. Assuming that the friction coefficient
is proportional to the number of ``regions'' in contact, $\mu(\ell)$ at
position $\ell$ takes the value
\begin{equation}
  \label{eq:mu}
  \mu(\ell) = b \frac{N(\ell)}{N_{\rm max}},
\end{equation}
where $N(\ell)$ is the number of microcontacts, $N_{\rm max}$ is the block
length in bits and $b$ is a constant that can be associated to the contact
stiffness.  Equation (\ref{eq:mu}) is similar to the one proposed by Bowden
and Tabor~\cite{tabor50}. The dynamic evolution dictated by equation
(\ref{eq:redk}) can be discretized and, after each displacement of length $a$
(one bit), the kinetic energy loss is calculated as
\begin{equation}
  \label{eq:deltae}
  \Delta k = a \left ( \tan(\theta) -  \frac{bN(t)}{N_{max}}\right).
\end{equation}
The block is moved on the plane in single-bit steps until the kinetic energy
vanishes. The critical angle $\theta_c$ is defined by taking $\av{\Delta k}=0$
in (\ref{eq:deltae}) and gives
\begin{equation}
  \label{eq:thetac}
  \tan\theta_c = {\overline \mu} = b C_p C_b,
\end{equation}
%
\begin{figure}[thpb]
  \centerline{\psfig{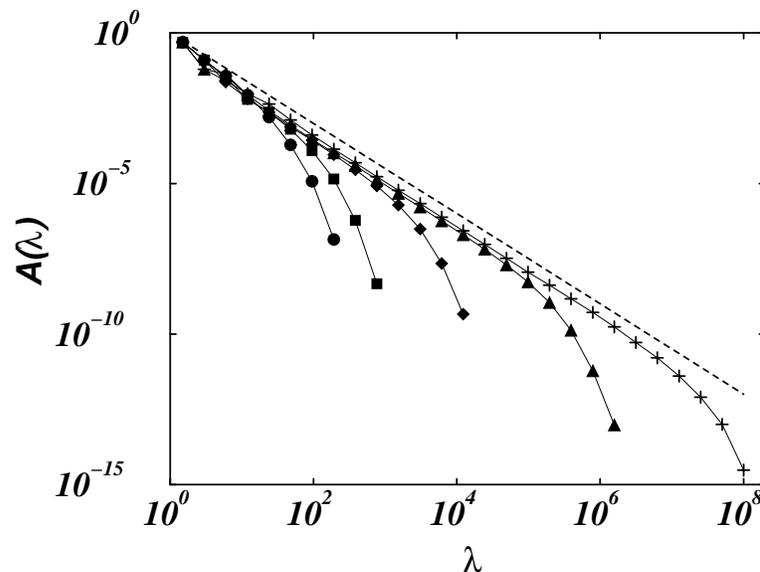}}
\caption{{}
  Distribution $A(\lambda)$ of stopping lengths as obtained numerically for
  the one-bit-block model. Averages were taken over $10^8$ realizations with
  an initial reduced kinetic energy $k_0=7.21~10^{-6} m$ ($v_0=10^{-2} m/s$ and
  $g=9.810 m/s^2$) and a critical angle $\theta_c=45^o$. The inclination angle
  $\theta$ of the plane was: $35^o$ (circles), $40^o$ (squares), $44^o$
  (diamonds), $44.9^o$ (triangles) and $44.99^o$ (crosses).  The dashed line
  corresponds to $A(\lambda)=\lambda^{-3/2}$. The same exponent was found
  experimentally \protect \cite{brito95}. }
\label{fig:2}
\end{figure}

In the limit in which the average sliding is much larger than the block size
in bits, (i.e. if $v_0$ is large, or $\theta$ is close to $\theta_c$) one does
not expect any dependence of the results on the length $N_{\rm max}$ of the
block, as long as the distribution of the friction coefficient $\mu$ has a
constant mean and width. In this case it is numerically convenient to take a
block length of one bit (which is always set to one). The plane bits on the
other hand are set to one with probability $C_b$. In this case $\mu$ takes the
values $0$ and $b$ with probabilities $1-C_p$ and $C_p$ respectively, so that
$\overline \mu=bC_b$ .  Fig.~\ref{fig:2} shows our numerical results for this
single-bit implementation. We have used $C_p=0.5$ and $b=2$, i.e. ${\overline
  \mu}=1$, therefore $\theta_c=\pi/4$.  The initial reduced kinetic energy was
$k_0=7.21~10^{-6} m$ ($v_0=10^{-2} m/s$ and $g=9.810 m/s^2$). Averages were
performed over $10^8$ realizations for each value of $\theta$. When $\theta
\to \theta_c$ we find that $A(\lambda) \sim \lambda^{-3/2}$, for $\lambda$
smaller than a $\theta$-dependent cutoff $\xi(\theta)$. This behavior is in
partial agreement with the experimental results of BG~\cite{brito95}. While
the exponent they find is consistent with $3/2$, they do not report any
evidence for the existence of a finite cutoff. According to our results, a
very large number of experimental realizations would be needed before a cutoff
can be clearly distinguished in $A(\lambda)$. As can be seen by integrating
the data in Fig. 2, for deviations from the critical angle as large as $10\%$,
$A(\lambda)$ only deviates from a power-law behavior for very large events,
which have a small probability $10^{-5}$ to happen. This means that one needs
of order $10^{5}$ realizations in order to assess the existence of a cutoff in
$A(\lambda)$. Notice however that BG only performed $10^3$ repetitions of
their measurements for each set of parameters, and this explains why only the
power-law regime is observed in their experiments.

\begin{figure}[thpb]
  \centerline{\psfig{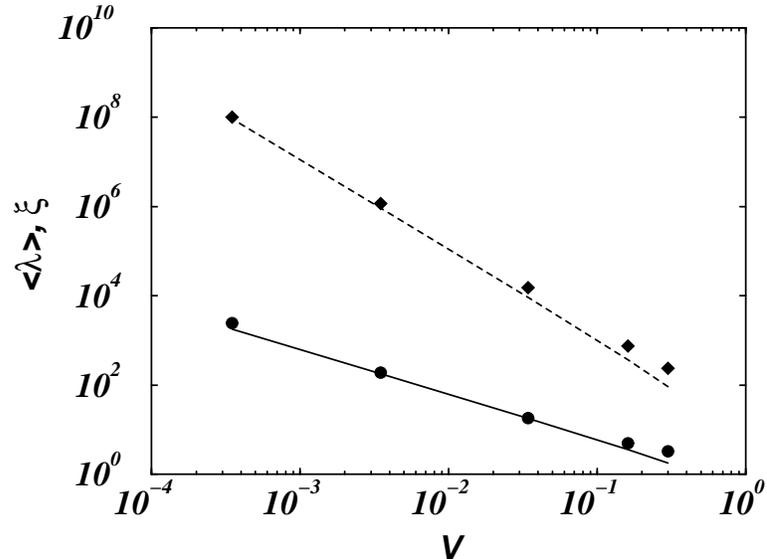}}
\caption{{}
  Numerical results for the mean stopping length (circles) and ``cutoff
  length'' (diamonds) as functions of $-V$. The solid line (dashed line)
  corresponds to $\av{\lambda} \propto |V|^{-1}$ ($\xi \propto |V|^{-2}$).}
\label{fig:3}
\end{figure}

Fig.~\ref{fig:3} displays the mean stopping length $\av{\lambda}$ and the
``cutoff'' $\xi$ versus $V \propto (\theta_c-\theta)$, calculated from the
data in Fig.~\ref{fig:2}.  When $\theta \rightarrow \theta_c$ ($V\to0$) we
find that $\av{\lambda} \sim (\theta_c-\theta)^{-1}$ and $\xi \sim
(\theta_c-\theta)^{-2}$ approximate well our numerical results. This is in
slight discrepancy with BG who report that $\av{\lambda} \sim
(\theta_c-\theta)^{-0.23}$~\cite{brito95}.

\section{Mapping to a FPT problem}
\label{sec:theory}

\begin{figure}[thpb]
  \centerline{\psfig{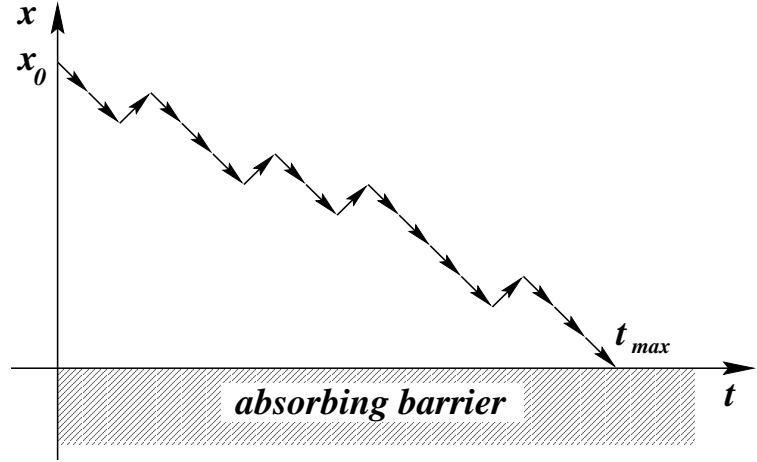}}
\caption{{} 
  A walker starts at position $x_0$ at $t=0$ and executes a random walk with
  mean drift $V<0$.  The distribution of times $t=t_{\rm max}$ for which the
  position becomes zero for the first time can be calculated from the
  probability distribution for a random walker with an absorbing barrier at
  $x=0$.  }
\label{fig:4}
\end{figure}

Since $k(\ell)$ satisfies equation~(\ref{eq:redk}) the problem of finding
$A(\lambda)$ is readily mapped onto a First-Passage-Time (FPT) problem for a
random walker with nonzero drift. The reduced kinetic energy $k(\ell)$ (which
is the ``position'' variable $x$ of the random walker), starts at
$x_0=k(0)=v_0^2/2g\cos{\theta}$, and executes a random walk with mean drift
$V=\tan{\theta} - \overline \mu$. In this picture $\ell$ has the meaning of a
``time'' variable, and we say that the sliding-block has stopped at time
$t_{\rm max}$ if its kinetic energy becomes zero at position $\lambda=t_{\rm
  max}$. Thus the distribution of stopping distances $A(\lambda)$ is the
distribution of First-Passage-Times for a random walker to cross $x=0$.  This
problem turns out to be exactly equivalent to the ``Gambler's Ruin''
problem~\cite{Feller,papoulis}, in which one asks for the probability for a gambler
with an initial capital $k_0$ not to have reached its ruin in $\lambda$ games
if it makes an average win $V$ in each run. Fig.~\ref{fig:4} shows a schematic
representation of the equivalent FPT problem.

Equivalently one can ask for the distribution probability $W(x,t)$ for a
random walker to be at position $x$ at time $t$, when there is an absorbing
barrier at $x=0$. The ``flux'' of particles at $x=0$ gives then the desired
distribution of First-Passage-Times $A(t)$. Because of these mappings, the
sliding-block problem with uncorrelated random friction turns out to be
completely equivalent to Compact Directed Percolation with an absorbing
wall~\cite{CDPW}(CDPW - See also~\cite{DK}), which is exactly solvable, and
analogous to Directed Percolation with an absorbing wall (DPW)~\cite{DPW},
which has not yet been solved analytically.

\begin{figure}[thpb]
  \centerline{\psfig{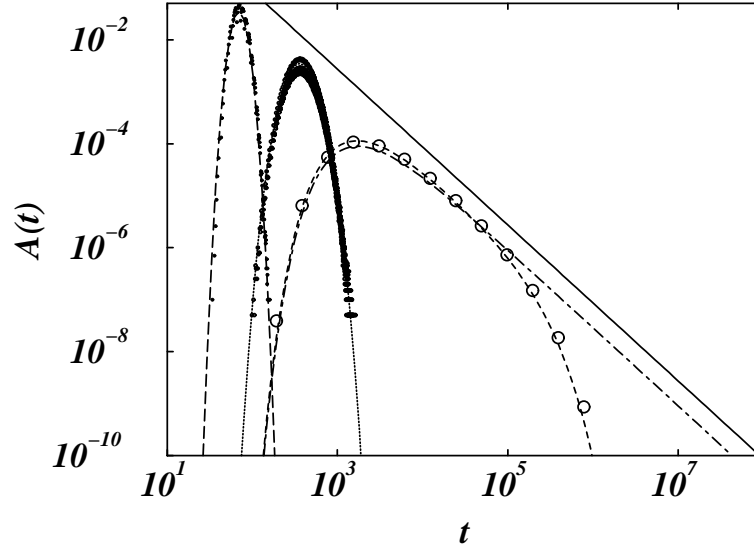}}
\caption{{}
  Comparison of numerical and analytical results for the stopping probability
  $A(t)$. Averages were taken over $10^7$ simulations with $k_0=7.21~10^{-2}m$
  ($v_0=1m/s$ and $g=9.810m/s^2$) for each chute inclination. Lines indicate
  the theoretical result (Equation~\ref{eq:pedete}) for $\theta=15^o$
  (long-dashed line), $40^o$ (dotted line), $44.9^o$ (dashed line) and
  $\theta=44.99^o$ (dot-dashed line).  For $\theta=15^o$ and $\theta=40^o$ we
  show the results of the simulation as small filled points.  The circles are
  the results for $\theta=44.9^o$. The solid line corresponds to the behavior
  found in the experiments \protect \cite{brito95}, $A(t) \propto t^{-3/2}$.}
\label{fig:7}
\end{figure}

Although this classical random-walk problem has been solved in many different
contexts~(e.g. \cite{Feller,papoulis,CDPW,DK}), an exact solution is briefly
derived here for self-containedness.  Let $W(x,t)$ be the probability for the
block to have reduced kinetic energy $k=x$ after traversing a distance
$\ell=t$. Since $k(\ell)$ satisfies the stochastic equation (\ref{eq:redk}),
$W(x,t)$ is a solution of the Fokker-Planck equation~\cite{risken}
\begin{equation}
\frac{\partial W(x,t)}{\partial t} = \left( -V 
\frac{\partial}{\partial x} + D
\frac{\partial^2}{\partial x^2} \right) W(x,t)
\label{eq:fp}
\end{equation} 
where $D=\sigma^2/2$. Since the particle stops, i.e., it is eliminated from
the system, when its kinetic energy becomes zero, one has to solve
(\ref{eq:fp}) with absorbing boundary condition at $x=0$
\begin{equation}
W(x,t) |_{x=0} = 0 \qquad \hbox{for all $t$}.
\label{eq:bc}
\end{equation}
The initial condition is $W(x,0)=\delta(x-k_0)$ if the block starts with a
well defined energy $k_0$. The Green function of (\ref{eq:fp}) is
\begin{equation}
G(x,t,V,D,k_0)=\frac{1}{\sqrt{4 \pi Dt}} \exp{\left\{-\frac{(x-k_0-Vt)^2}{4Dt}\right\}}
\end{equation}
in terms of which the solution of (\ref{eq:fp}) plus (\ref{eq:bc}) is\cite{fpt}
\begin{equation}
W(x,t)=G(x,t,V,D,k_0)-  G(x,t,V,D,-k_0) \quad e^{-k_0V/D}
\label{eq:solution}
\end{equation}
The probability $P(t)$ for the block not to have stopped (the random walker
not to have been absorbed) at time $t$ is
\begin{equation}
P(t)=\int_{0}^\infty W(x,t) dx
\end{equation}
and thus the probability $A(t)$ to be absorbed at time $t$ is
\begin{equation}
A(t)= - \frac{\partial P(t) }{\partial t} = -\int_{0}^\infty \dot{W}(x,t) dx
\end{equation}
Now using the fact that $W(x,t)$ satisfies the Fokker-Planck equation
(\ref{eq:fp}), it is readily shown that
\begin{equation}
A(t)= - S(x,t) |_{x=0}
\end{equation}
where $S(x,t)$ is the conserved \emph{flux}
\begin{equation}
S(x,t) = \left(V - D \frac{\partial }{\partial x} \right) W(x,t)
\end{equation}
so that finally
\begin{equation}
A(t) = \frac{k_0}{\sqrt{4 \pi Dt^3}} \exp{\left\{-\frac{(k_0+Vt)^2}{4Dt}\right\}}
\label{eq:pedete}
\end{equation}
In Fig.~\ref{fig:7} we compare this exact solution with our numerical
measurements for the single-bit model.  For a RW with step length $a=1$ is
readily found that $D=0.5$. We set $k_0=7.21~10^{-2}m$ ($v_0=1m/s$ and
$g=9.810m/s^2$). The agreement between analytical and numerical results is
very good.

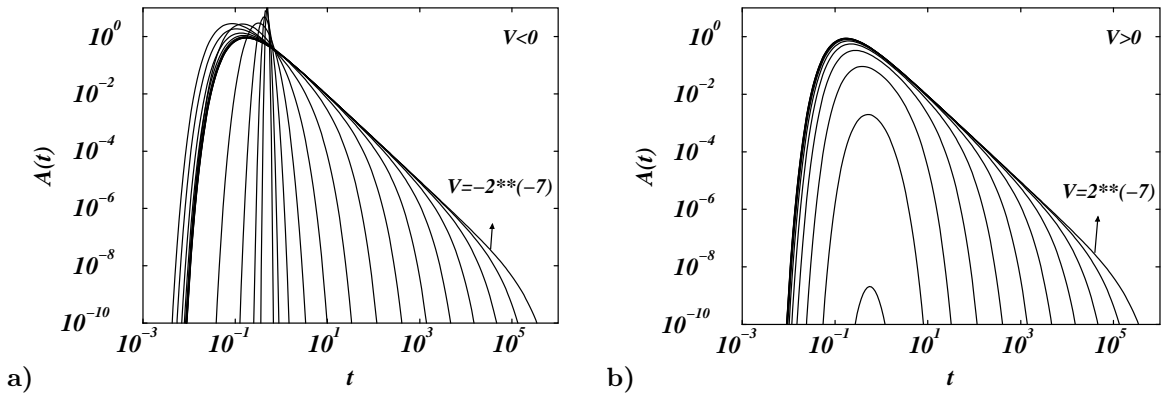
\begin{figure}[thpb]
  \centerline{{\bf a)}\psfig{figure=fign5a.epsi,width=7cm,angle=270} \hskip
    0.5cm {\bf b)}\psfig{figure=fign5b.epsi,width=7cm,angle=270}}
\caption{{}
  Stopping probability per unit time $A(t)$ for $D=0.5$, $k_0=7.21~10^{-4}m$
  ($v_0=10{-1} m/s$ and $g=9.81~m/s^2$). The drift $V$ takes the values {\bf a)} $-(1/2)^{7}, -(1/2)^{6},
  -(1/2)^{5},...$, and {\bf b)} $(1/2)^{7}, (1/2)^{6}, (1/2)^{5},...$.  For
  $V=0$ one has that $A(t) \sim t^{-3/2}$.}
\label{fig:5}
\end{figure}

Fig.~\ref{fig:5} shows $A(t)$ from equation (\ref{eq:pedete}) for several
values of $V$ which are taken to be powers of $1/2$ for convenience. Notice
that for $V<0$ the area under $A(t)$ is constant and equal to one, meaning
that the block always stops.  For positive $V$ on the other hand, this area is
less than one, meaning that there is a finite ($V$-dependent) probability
for the block never to stop, i.e. to ``escape'' to infinity.

Exactly at the critical angle (i.e. for $V=0$) one obtains
\begin{equation}
A(t) = \frac{k_0}{\sqrt{4 \pi Dt^3}} \exp{\left(-\frac{k_0^2}{4Dt}\right)}
\end{equation}
For large times ($Dt>>k_0^2$) this gives
\begin{equation}
A(t) \sim \frac{k_0}{\sqrt{4 \pi Dt^3}},
\label{eq:critical}
\end{equation}
which is consistent with our numerical measurements.

\begin{figure}[thpb]
  \centerline{\psfig{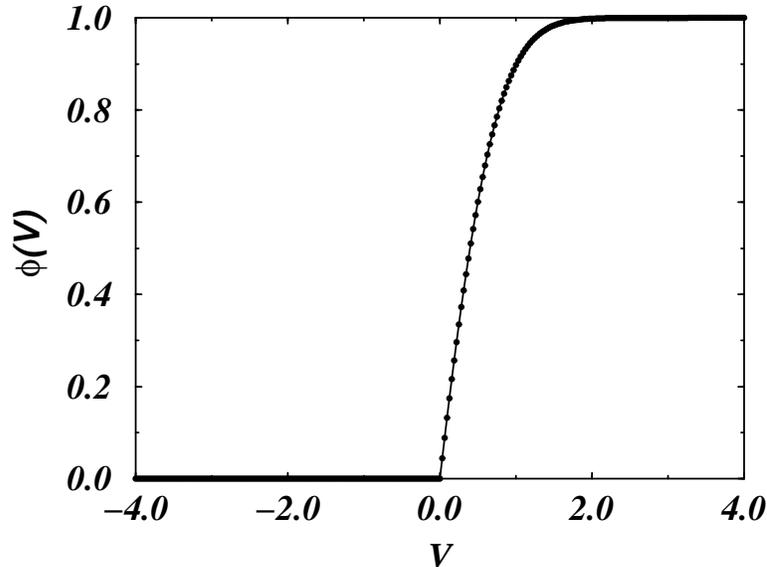}}
\caption{{}
  Order parameter $\phi = P(\infty) = 1 - \int_0^\infty A(\tau) d\tau$ as
  function of $V$. As can be seen $\phi=0$ when $V<0$, i.e. the block always
  stops, which is a consequence of the drift pushing it towards the barrier
  (the angle being less than critical). For positive $V$ on the other hand,
  the drift tends to push the particle away from the barrier (the angle is
  larger than critical) and $\phi>0$, i.e.  there is a finite probability of
  escape to infinity.  There is a second-order phase transition at $V_c=0$.}
\label{fig:6}
\end{figure}

The escape probability $\phi=P(\infty)$ is plotted in Fig.~\ref{fig:6} as a
function of $V$.  This probability is small if $V$ is small, thus there is a
continuous phase transition at $V_c=0$. As customary~\cite{CDPW}, for $V \sim
0+$ we write
\begin{equation}
\phi(V) \sim V^{\beta_1}, 
\label{eq:pinf} 
\end{equation}
which defines the critical exponent $\beta_1$. For finite times, $P(t,V) = 1 -
\int_0^t A(\tau) d\tau$ measures the probability for the particle to be
``alive''. Usual scaling arguments allow one to write, for $t$ large and
$|V|<<1$,
\begin{equation}
P(t,V) \sim t^{-\delta} f(t/\xi(V))
\label{eq:P(t)}
\end{equation}
with $\xi(V)$ a correlation time diverging at $V_c=0$ as $\xi \sim
|V|^{-\nu_{\parallel}}$, and $\delta = \beta_1/\nu_{\parallel}$. The scaling
function $f(x)$ satisfies $f \to const.$ when $x\to 0$, thus when $V=0$ one
has that $P_t \sim t^{-\delta}$, i.e. the power-law decay of correlations that
is typical of a critical point.

Now it is easy to calculate $\beta_1$ and $\nu_{\parallel}$. Since $\partial
P(t,V)/\partial t=-A(t)$ one has that at $V=0$ $A(t)$ behaves as
$t^{-(1+\delta)}$. Therefore equation (\ref{eq:critical}), implies
$\delta=1/2$, in agreement with BG experimental results \cite{brito95}.

The ``cutoff'' time $\xi$ for finite but small $V$ results from the condition
that the argument of the exponential in (\ref{eq:pedete}) be larger than one.
Thus solving $(k_0 + V \xi)^2= 4 D \xi$ one obtains $\xi\sim 2D/V^2$ i.e.
$\nu_{\parallel}=2$.  Therefore $\beta_1=1$.  This last value can be confirmed
using (\ref{eq:pinf}), since
\begin{eqnarray}
\phi(V) &=& 1 - \int_0^\infty  A(\tau,V) d\tau =  \int_0^\infty  \{
A(\tau,-V)-A(\tau,V)\} d\tau  \nonumber\\ 
&=& 2 \sinh\left(\frac{V k_0}{2D}\right) \int_0^\infty 
\frac{k_0}{\sqrt{4 \pi Dt^3}} \exp{\left(-\frac{k_0^2+ V^2t^2}{4Dt}\right)} 
\end{eqnarray}
which for small $V$ gives $\phi \sim V$ since the integral gives a
constant value in this limit. 

The third independent exponent, and the last one needed to fully characterize
the critical behavior in DP is the ``meandering exponent'' $\chi$ defined by
$\av{x^2} - \av{x}^2 \sim t^{\chi}$ with $\chi = 2
\nu_{\perp}/\nu_{\parallel}$ and $\nu_{\perp}$ associated to the divergence of
``space'' correlations $\xi_{\perp} \sim |V|^{-\nu_{\perp}}$. For a random
walk we have $\av{x^2}-\av{x}^2 \sim t$ and thus $\chi=1$ implying
$\nu_{\perp}=1$.

From the values of these exponents one can conclude that for $V \to 0^-$ the
mean stopping time $\av{\lambda}$ behaves as $\av{\lambda} \sim |V|^{-\tau_1}$
with~\cite{DPW} $\tau_1 = \nu_{\parallel} - \beta_1=1$. This again is in good
agreement with our numerical measurements.

\section{Conclusions}
\label{sec:conclusions}
This work shows that most of the experimental results obtained by Brito \&
Gomes for sliding blocks on a chute~\cite{brito95} can be reproduced by a very
simple model. Compared with the traditional problem of a block sliding on a
chute, a random friction coefficient is the only new ingredient in our study.
The problem of finding the distribution of stopping lengths is equivalent to a
first-passage-time random walk problem for an uncorrelated random walker with
zero drift, and thus exact analytical solution. We derive this solution and
compare its predictions with numerical results, obtaining a perfect agreement.
At the critical angle $\theta_c=\arctan \overline \mu$, a power-law
distribution of stopping distances is obtained: $A(\lambda) \sim
\lambda^{-3/2}$ in good agreement with experimental findings. However, a
discrepancy arises for the mean sliding length $\av{\lambda}$, which is in
this work found to behave as $\av{\lambda} \sim |V|^{-1}$, while BG report
$\av{\lambda} \approx |V|^{-0.23}$. We believe that this difference is due to
uncontrolled experimental errors, mainly because of the difficulty involved in
the measurement of $\av{\mu}$ (and thus $\theta_c$) on real systems.

\section*{Acknowledgments}
C.M. wishes to thank Kent Lauritsen for useful discussions on Directed
Percolation, and also for pointing out the equivalence between our model and
Compact Directed Percolation with a wall.  The authors acknowledge financial
support from Brazilian agencies FAPERJ, CNPq and CAPES.

\end{document}